\documentstyle[12pt]{article}

\newcommand{\sect}[1]{\setcounter{equation}{0}\section{#1}}

\newcommand{\eq}{\begin{equation}}
\newcommand{\eqa}{\begin{eqnarray}}
\newcommand{\en}{\end{equation}}
\newcommand{\ena}{\end{eqnarray}}
\newcommand{\enn}{\nonumber \end{equation}}

\def\sk{\vskip .4cm}
\def\noi{\noindent}
\def\om{\omega}

\def\Ga{\Gamma}

\let \si\sigma
\let \part\partial

\def\unmezzo{{1 \over 2}}
\def\epsi{\varepsilon}
\def\we{\wedge}

\def\de{\delta}

\def\tv{{\bf t}}
\def\vbo{{\bf v}}
\def\Gt{{\tilde G}}

\def\part{\partial}

\def\sk{\vskip .4cm}

\def\noi{\noindent}

\def\X0{X^0}

\def\om{\omega}

\def\unmezzo{{1 \over 2}}
\def\epsi{\varepsilon}
\def\epsibo{{\bf  \epsilon}}

\def\psib{{\bar \psi}}

\def\we{\wedge}

\def\de{\delta}
\def\CABC{{C^A}_{BC}}

\def\tbo{{\bf t}}
\def\Gt{{\tilde G}}

\def\c#1#2{ C_{~#1}^{#2} }
\def\cl#1#2{ C_{~#1}^{#2} }

\def\square{{\,\lower0.9pt\vbox{\hrule \hbox{\vrule height 0.2 cm
\hskip 0.2 cm \vrule height 0.2 cm}\hrule}\,}}

\begin{document}

\begin{titlepage}
\rightline{DFTT-52/95}
\rightline{hep-th/9509031}
\rightline{August 1995}
\vskip 2em
\begin{center}{\bf   FREE DIFFERENTIAL ALGEBRAS: THEIR
USE IN FIELD THEORY AND DUAL FORMULATION}
\\[3em]
Leonardo Castellani\\[1em]
{\sl II Facolt\`a di Scienze M.F.N. di Torino, sede di
Alessandria}\\[.5 em]
{\sl Dipartimento di Fisica Teorica\\
and\\Istituto Nazionale di
Fisica Nucleare\\
Via P. Giuria 1, 10125 Torino, Italy.}  \\[3em]
Alberto Perotto\\[1em]
{\sl Dipartimento di Fisica Teorica\\
Via P. Giuria 1, 10125 Torino, Italy.}  \\[3em]

\end{center}

\begin{abstract}
The gauging of free differential algebras (FDA's)
produces gauge
field theories containing antisymmetric tensors.
The FDA's extend the Cartan-Maurer equations of ordinary
Lie algebras by incorporating  $p$-form potentials ($p > 1$).
We study here the algebra of FDA transformations.
To every $p$-form in the FDA we associate an extended
Lie derivative $\ell$ generating a corresponding ``gauge"
transformation. The field theory based on the FDA
is invariant under these new transformations.
This gives geometrical meaning to the antisymmetric
tensors. The algebra of Lie derivatives is shown to close
and provides
the dual formulation of FDA's.
\end{abstract}

\vskip .2cm
\noi \hrule
\vskip.2cm
\noi {\small{\it e-mail:} castellani@to.infn.it}
\vskip 1cm
\centerline{Talk presented by. L. Castellani at the $4^{th}$
Colloquium}
\centerline{``Quantum groups and integrable systems", Prague, June
1995.}
\end{titlepage}

\newpage
\setcounter{page}{1}

\sect{Introduction}

Free differential algebras \cite{fda1,fda2,gm2} have emerged
as underlying symmetries of
field theories containing antisymmetric tensors, as for example
supergravity and superstring theories. Within the group-geometric
method of ref.s \cite{gm1,gm2,gm3}, a systematic algorithm exists
that
produces lagrangians invariant under any given FDA.
\sk
Actually, the
FDA symmetries related to the antisymmetric tensors were not
explicitly discussed in \cite{fda1,fda2,gm2}. They were treated
in \cite{fda3,gm3}, where they could be deduced from the BRST
algebra of FDA's by interpreting the ghosts as gauge parameters.
\sk
Here we present a direct geometric interpretation
of antisymmetric tensors: they are the gauge fields
of gauge transformations generated by a new type of Lie
derivative. These new Lie derivatives, together with the
usual Lie derivatives along the Lie algebra tangent vectors,
close on an algebra that can be called the dual formulation
of FDA's, extending ordinary Lie algebras.
\sk
Our arguments are developed in the case of the simplest
FDA extension of the Cartan-Maurer equations,  containing a 2-form.

\sect{Group geometry and dynamical fields}

We sketch the basic steps of the group-geometric approach
of ref.s \cite{gm1,gm2}. See \cite{gm3} for a short review.
\sk
\noi {\sl Lie algebra, tangent vectors, vielbeins}
\sk
Consider an ordinary
Lie algebra Lie({\sl G}), with abstract generators $T_A$
satisfying the commutation relations
\eq
[T_A,T_B]={C^C}_{AB}T_C  \label{Lie}
\en
On the group manifold $G$ we can find a basis of tangent vectors
${\bf t}_A$ closing on the same algebra as in (\ref{Lie}). Their
duals are the left-invariant one-forms $\si^A$ (cotangent
basis), also called vielbeins, satisfying
the Cartan-Maurer equations:
\eq
d\si^A+{1 \over 2}{C^A}_{BC}\si^B \we \si^C = 0 \label{CM}
\en
as can be seen by using (\ref{Lie}) and $\si^A (\tv_B) = \de_A^B$.
Thus, the commutation algebra (\ref{Lie}) and the Cartan-Maurer
equations (\ref{CM}) are equivalent descriptions of the same
group structure. The Jacobi identities
\eq
C^A_{~~B[C} C^B_{~~DE]} =0  \label{Jacobi}
\en
necessary for the consistency of (\ref{Lie}) ensure the
integrability of eq.s (\ref{CM}), that is the nilpotency
of the external derivative $d^2=0$.
\sk
\noi {\sl Dynamical fields, curvatures and Bianchi identities}
\sk
The main idea of ref.s \cite{gm1,gm2} is to consider the
one-forms $\si^A$ as the fundamental fields of the
geometric theory to be constructed. More precisely, the
dynamical fields are the vielbeins $\mu^A$ of
$\Gt$, a smooth deformation of the group manifold $G$
referred to as ``soft group manifold".
In general $\mu^A$ does not satisfy the Cartan-Maurer
equations any more, so that
\eq
d\mu^A+{1 \over 2} \CABC \mu^B \we \mu^C \equiv R^A \not= 0 \label{R}
\en

\noi The extent of the deformation $G \rightarrow \Gt$ is measured by
the curvature two-form $R^A$. $R^A = 0$ implies $\mu^A=\si^A$ and
viceversa. The deformation
is necessary in order to allow configurations with
nonvanishing curvature.
\sk
Applying the external derivative $d$ to the definition (\ref{R}),
using $d^2
=0$ and the Jacobi identities (\ref{Jacobi}), yields
the Bianchi identities
\eq
(\nabla R)^A \equiv dR^A - \CABC R^B \we \mu^C =0 \label{Bianchi}
\en
\sk
\noi {\sl An example: $G$ = Poincar\'e group}
\sk
Consider $\Gt$= smooth
deformation of the Poincar\'e
group, whose structure constants are read off the corresponding
Lie algebra :

\eqa
& &[P_a, P_b]=0\\
& &[M_{ab}, M_{cd}]=\eta_{ad}M_{bc}+\eta_{bc}M_{ad}-
\eta_{ac}M_{bd}-\eta_{bd} M_{ac} \\
& &[M_{ab}, P_c]=\eta_{bc} P_a-\eta_{ac} P_b \label{LiePoincare}
\ena

Denoting by $V^a$ and $\om^{ab}$ the vielbein $\mu^A$ when the index
A runs on the
translations and on the Lorentz rotations
respectively, eq.s (\ref{R}) take
the form:

\eqa
& &R^a=d V^a- \om^{ab} \we V^c \eta_{bc} \\
& &R^{ab}=d\om^{ab}-\om^{ac} \we \om^{db} \eta_{cd} \label{RPoincare}
\ena

The fundamental fields $V^a$ and $\om^{ab}$ are interpreted as the
ordinary vierbein and the spin connection, respectively, and
 eq.s (\ref{RPoincare})
define the torsion and the Riemann curvature. These satisfy
the Bianchi
identities
\eqa
& &dR^a-R^{ab}V^b+\om^{ab} R^b \equiv {\cal D}R^a-R^{ab}V^b=0 \\
& &dR^{ab}-R^{ac} \om^{cb}+\om^{ac}R^{cb} \equiv {\cal D} R^{ab} =0
   \label{BianchiPoincare}
\ena
\noi Products between forms are understood to be
exterior products,
${\cal D}$ is the Lorentz covariant derivative, and repeated
indices are contracted with the Minkowski metric $\eta_{ab}$.
\sk
Note that the fields $\mu^A(y)$ depend on
all the soft group manifold coordinates $y$. In the Poincar\'e
example,
this means that the vierbein and the spin connection depend on the
coordinates $y^a$ associated to the translations (the ordinary space-
time coordinates) {\sl and} on the coordinates $y^{ab}$ associated to
the Lorentz rotations. Since we want to have {\sl space-time} fields
at the end of the game, we have to find a way to remove the $y^{ab}$
dependence. This is achieved when the curvatures are {\sl horizontal}
in
the $y^{ab}$ directions (see later).
\sk
How do we find the dynamics of $\mu^A (y)$ ? We want to obtain a
geometric theory, i.e. invariant under diffeomorphisms of the soft
group
manifold $\Gt$. We need therefore to construct an action invariant
under
diffeomorphisms, and this is simply achieved by using only
diffeomorphic
invariant operations as the exterior derivative and the wedge
product.
The building blocks are the one-form $\mu^A$ and its curvature
two-form
$R^A$, and exterior products of them can make up a lagrangian D-form
(where D is the dimension of space-time).
\sk
\noi {\sl Diffeomorphisms}
\sk
The variation under diffeomorphisms $y+\epsi$ of the vielbein
field $\mu^A(y)$ is given by the Lie derivative of the vielbein
along the infinitesimal tangent vector $\epsibo \equiv \epsi^A
\tbo_A$:

\eq
\de \mu^A = \mu^A (y+\epsi)-\mu^A(y)=d(i_{\epsilon}\mu^A)
       +i_{\epsilon} d \mu^A \equiv \ell_{\epsilon} \mu^A
\label{diff}
\en

On p-forms $\om_{(p)}=\om_{B_1...B_p}\mu^{B_1} \we ... \we \mu^{B_p}
$,
the {\sl contraction} $i_{\bf v}$ along an arbitrary tangent vector
${\bf v}=
v^A  {\bf t}_A$
is defined as
\eq
i_{\bf v}~ \om_{(p)}=p~ v^A \om_{AB_2...B_p}~\mu^{B_2}
 \we ... \we \mu^{B_p}
    \label{contraction}
\en
and maps $p$-forms into ($p-1$)-forms.

The operator
\eq
l_{\vbo} \equiv d~ i_{\vbo} + i_{\vbo} ~d  \label{Liederivative}
\en
\noi is the {\sl Lie derivative} along the tangent vector $\vbo$
and maps $p$-forms into $p$-forms.
Eq. (\ref{diff}) gives the variation under diffeomorphisms of
any $p$-form.

\sk
We now rewrite the variation $\de \mu^A$ of eq. (\ref{diff}) in a
suggestive
way, by adding and subtracting $\CABC \mu^B  \epsi^C$ :

\eqa
& & \de \mu^A = d \epsi^A + \CABC \mu^B  \epsi^C - 2 \mu^B \epsi^C
(d\mu^A)_{BC} - \CABC \mu^B \epsi^C  \nonumber \\
& &= (\nabla \epsi)^A + i_{\epsibo} R^A
      \label{diffbis}
\ena

\noi where we have used the definition (\ref{R}) for the curvature,
and
the $G$-covariant derivative $\nabla$ acts on $\epsi^A$ as
\eq
(\nabla \epsi)^A \equiv d \epsi^A + \CABC \mu^B \epsi^A
\label{covdev}
\en
\sk
\noi {\sl Horizontality}
\sk
All the invariances of the geometric theory are contained in eq.
(\ref{diffbis}).
In particular, suppose that the two-form $R^A={R^A}_{BC} ~\mu^B \we
\mu^C$
has vanishing components along the directions of a subgroup $H$ of
$G$:
\eq
{R^A}_{BH}=0~~~~~ \vbox{\hbox {\rm \small A~runs~on~G} \hbox{\rm
\small H~runs~on~H}} \label{horizontality}
\en
\noi Then we say that $R^A$ is {\sl horizontal} on $H$, and the
diffeomorphisms along the $H$-directions reduce to {\sl gauge
transformations}:
\eq
\de \mu^A(y)=(\nabla \epsi)^A \label{gaugetransf}
\en
Moreover, the dependence on the $y^H$ coordinates becomes
inessential,
in the sense that it factorizes after a finite gauge transformation
(see ref.s \cite{gm2,gm3}).
The theory "remembers" the invariance under $y^H$-diffeomorphisms by
retaining the gauge invariance under $H$, with $\epsi^H$
interpreted now as a gauge parameter.
\sk
For example, in Poincar\'e gravity  the curvatures are
horizontal along
the Lorentz directions :
then the fields $V^a$ and $\om^{ab}$ live on the
coset space
\eq
{\rm {G \over H}={Poincare' \over Lorentz}}
\en
\noi i.e. on ordinary spacetime. The lagrangian is integrated on a D-
volume (D-dimensional spacetime), and is therefore a D-form. The
resulting theory is invariant
under D-spacetime diffeomorphisms, and under local Lorentz rotations.
\sk
Finally, we recall the algebra of Lie derivatives on the group
manifold.
{}From
\eq
\ell_{ \epsi^B \tbo_B}\mu^A = d\epsi^A + \left( C^A_{BC} -
2R^A_{BC}\right)
\mu^B\epsi^C \label{Liemu}
\en
\noi cf.  (\ref{diffbis}), we deduce
\eq
\left[ \ell_{ \epsi^A_1 \tbo_A},\ell_{ \epsi^B_2 \tbo_B} \right] =
\ell_{
\left[
\epsi^A_1 \partial_A \epsi^C_2 - \epsi^A_2 \partial_A \epsi^C_1
+\epsi^A_1 \epsi^B_2 \left( C^C_{AB} -2R^C_{AB}\right) \right]\tbo_C
}
\label{Liemualgebra}
\en
\noi where the partial derivative $\part_A $ of a function $f$ is
defined by
$df \equiv (\part_A f) \mu^A$ or also $\part_A f \equiv \tv_A (f)$.
One can
can verify the standard formula:
\eq
[\ell_{\epsi_1^A \tbo_A},\ell_{\epsi_2^B \tbo_B}]=\ell_{[\epsi_1^A
\tbo_A,~
\epsi_2^B \tbo_B]} \label{Lieclosuremu}
\en
In particular,  the Lie derivatives $\ell_{\tv_A}$ on the undeformed
group
manifold $G$ close on the Lie algebra:
\eq
[\ell_{\tv_A}, \ell_{\tv_B}]=C^C_{~AB} \ell_{\tv_C} \label{rigidLie}
\en

\sect{Free differential algebras}
\sk
The dual formulation of Lie algebras provided by the Cartan-Maurer
equations (\ref{CM}) can be naturally extended to $p$-forms ($p >1$):

\eq
d \theta^i_{(p)}+\sum {1 \over n} C^i_{~i_1...i_n} \theta^{i_1}_{(p_
1)}\we ...\we \theta^{i_n}_{(p_n)}=0,~~~~~p+1=p_1+...+p_n
\label{FDA}
\en

\noi $p, p_1,...p_n$ are the degrees of the forms $\theta
^i, \theta^{i_1},..., \theta^{i_n}$; the indices
$i, i_1,..., i_n$ run on
irreps of a group $G$, and $C^i_{~i_1...i_n}$ are generalized
structure
constants satisfying generalized Jacobi identities due to $d^2=0$.
When
$p=p_1=p_2=1$ and $i, i_1, i_2$ belong to the adjoint representation
of
$G$, eq.s (6.1) reduce to the ordinary Cartan-Maurer equations. The
(anti)symmetry properties of the indices $i_1,...i_n$ depend on the
bosonic or fermionic character of the forms $\theta^{i_1},...\theta^
{i_n}$

If the generalized Jacobi identities hold, eq.s (\ref{FDA})
define a {\sl free
differential algebra} (FDA).
The possible FDA extensions $G'$ of a Lie
algebra $G$ have been studied in ref.s  \cite{fda1,fda2}, and rely
on the existence of
Chevalley cohomology classes in $G$ \cite{Che}. Suppose that, given
an ordinary
Lie algebra $G$, there exists a $p$-form:
\eq
\Omega^i_{~(p)} (\sigma)=\Omega^i_{~A_1...A_p} \sigma^{A_1} \we
...\we
\sigma^{A_p},~~{\rm \Omega^i_{~A_1...A_p} =constants,~i~runs~on~a~G-
irrep} \label{Omega}
\en

\noi which is covariantly closed but not covariantly exact, i.e.
\eq
\nabla \Omega^i_{~(p)} \equiv d\Omega^i_{~(p)} + \sigma^A \we D(T_A)^
i_{~j} \Omega^j_{~(p)}=0,~~~~\Omega^i_{~(p)} \not= \nabla
\Phi^i_{~(p-1)}
\label{coclass}
\en

\noi Then $\Omega^i_{~(p)}$ is said to be a representative of a
Chevalley cohomology class in the $D^i_{~j}$ irrep of $G$. $\nabla$
is the boundary operator satisfying $\nabla^2=0$ (it would be
proportional to the curvature 2-form on the {\sl soft} group
manifold).
The existence of $\Omega^i_{~(p)}$ allows the extension of the
original
Lie algebra $G$ to the FDA $G'$:

\eqa
& &d \sigma^A+\unmezzo \CABC \sigma^B \we \sigma^C=0 \label{FDALie}\\
& &\nabla \Sigma^i_{~(p-1)} + \Omega^i_{(p)} (\sigma)=0
\label{FDAnew}
\ena

\noi where $\Sigma^i_{(p-1)}$ is a new $(p-1)$-form, not contained in
$G$.
Closure of eq.s (6.4) is due to $\nabla \Omega^i_{(p)} = 0$.

It is clear that $\Omega^i_{(p)}$ differing by exact pieces $\nabla
\Phi
^i_{(p-1)}$ lead to equivalent FDA's, via the redefinition
$\Sigma^i_{(p
-1)} \rightarrow \Sigma^i_{(p-1)} + \Phi^i_{(p-1)}$. What we are
interested in are really nontrivial cohomology classes satisfying
eq.s
(\ref{coclass}).

The whole game can be repeated on the free differential algebra $G'$
which now contains $\sigma^A$, $\Sigma^i_{(p-1)}$. One looks for the
existence of polynomials in $\sigma^A$, $\Sigma^i_{(p-1)}$
\eq
\Omega^i_{(q)} (\sigma, \Sigma)= \Omega^i_{A_1...A_r i_1...i_s}
\sigma
^{A_1} \we ... \we \sigma^{A_r} \we \Sigma^{i_1}_{(p-1)} \we ... \we
\Sigma^{i_s}_{(p-1)}
\en

\noi satisfying the cohomology conditions (\ref{coclass}).
If such a polynomial
exists, the FDA of eq.s (\ref{FDALie}), (\ref{FDAnew})
can be further extended to $G''$, and so
on.

\noi {\sl Note 1}: In constructing $D$-dimensional supergravity
theories we usually choose
as starting point the superPoincar\'e Lie algebra.
The possible $G'$ extensions to
FDA's depend on the spacetime dimension $D$. For
example in $D=1 1$ there
is a cohomology class of the superPoincar\'e algebra in the identity
representation:

\eq
\Omega (V,\om,\psi)=\unmezzo \psib  \Ga^{ab} \psi  V^a  V^b
\en

\noi where $\psi$ is the gravitino field, dual to
the supersymmetry charge;  $d\Omega=0$ holds because of the $D=11$
Fierz
identity

\eq
\psib  \Ga^{ab} \psi ~ \psib  \Ga^a \psi V^b=0
\en

\noi This allows the extension of the algebra (3.4) by means of a
three
-form $A$:
\eq
dA-\Omega (V,\om,\psi)=0
\en
\noi {\sl Note 2:} Only nonsemisimple algebras can have FDA
extensions in
nontrivial $G$-irreps. Indeed a theorem by Chevalley and Eilenberg
\cite{Che}
states that there is no nontrivial cohomology class of $G$ in
nontrivial $G$-irreps when $G$ is semisimple.
\sk
As we have done in the case of ordinary Lie algebras, we find a
dynamical theory based on FDA's by allowing nonvanishing curvatures.
This means, for example, that $D=11$ supergravity is based on a
deformation of the fields $V,\om,\psi,A$ such that the
superPoincar\'e
curvatures and the $A$-curvature are different from zero.
For the geometric construction of the action
we refer the reader to refs. \cite{gm2,gm3}.  Other theories
with antisymmetric tensors have been
interpreted as gaugings of free differential algebras: see
\cite{gm2} for a detailed study.
\sk
\sect{The FDA1 algebra}

We consider here the simplest extension of a Lie algebra,
denoted by FDA1:

\eqa
& & d \sigma^A + \unmezzo \CABC ~\sigma^B \sigma^C = 0
\label{fdaLie}\\
& & dB^i+C^i_{~Aj} \sigma^A B^j + {1 \over 6} C^i_{~ABC} \sigma^A
\sigma^B \sigma^C \equiv \nabla B^i + {1 \over 6} C^i_{~ABC} \sigma^A
\sigma^B \sigma^C =0 \label{fdaB}
\ena

\noi where $B^i$ is a two-form in a representation $D^i_{~j}$ of $G$.
The generalized Jacobi identities ($d^2=0$), besides the usual ones
for
$\CABC$, are

\eq
C^i_{~Aj} C^j_{~Bk} - C^i_{~Bj} C^j_{~Ak} = C^C_{~AB} C^i_{~Ck},
{}~~~{\rm representation~condition}  \label{jacobi2}
\en
\eq
4 C^j_{~[ABC} C^i_{~D]j} + 6 C^E_{~[AB} C^i_{~CD]
E}=0,~~~{\rm 3-cocycle~condition} \label{jacobi3}
\en

\noi Eq. (\ref{jacobi2}) implies that $(C_A)^i_{~j} \equiv C^i_{~Aj}$
is a
matrix
representation of $G$, while eq. (\ref{jacobi3}) is just the
statement that
$C^i
\equiv C^i_{~ABC} \sigma^A \sigma^B \sigma^C$ is a 3-cocycle, i.e.
$\nabla C^i = 0$.
If we allow
the left hand sides of eq.s (\ref{fdaLie}), (\ref{fdaB})
to be nonvanishing curvatures $R^A$,
$R^i$ respectively, we find the  Bianchi identities:
\eqa
       & & dR^A-\CABC~ R^B \mu^C=0 \\
       & & dR^i-C^i_{~Aj} R^A B^j+C^i_{~Aj} \mu^A R^j - {1 \over 2}
        C^i_{~ABC} R^A \mu^B \mu^C=0
\ena
\noi where we use the same symbol $B^i$ for the "soft" 2-form. The
curvatures can be expanded on the $\mu^A, B^i$ basis
as
\eqa
& &R^i=R^i_{~ABC} \mu^A \mu^B \mu^C + R^i_{~Aj} \mu^A B^j
\label{Bianchi1}\\
& &R^A={R^A}_{BC} \mu^B \mu^C+R^A_{~i } B^i \label{Bianchi2}
\ena
\sk
\noi {\sl Lie derivatives}
\sk
Using the definition (\ref{Liederivative}) and the expression of the
FDA curvatures, we find the action of the Lie derivative on
$B^i$:
\eqa
& &\ell_{ \epsi^Bt_B}B^i = \left(R^i_{Aj} - C^i_{Aj}\right)\epsi^AB^j
+
\left(3R^i_{ABC} -\frac{1}{2}C^i_{ABC}\right)\epsi^A\mu^B\wedge\mu^C
\label{LieB}
\ena
\noi the action on $\mu^A$ remaining the one given in (\ref{Liemu}).
Here we find something interesting: the Lie derivatives
do not close any more as in eq. (\ref{Liemualgebra}). Before
computing this modified algebra, let us define
\sk
i) a new contraction operator $i_{\epsi^j \tv_j}$ by its action on a
generic
p-form $\om = \om_{i_1...i_n A_1...A_m} B^{i_1} \we ... B^{i_n} \we
\mu^{A_1}
\we ... \mu^{A_m}$ as
\eq
i_{\epsi^j \tv_j} \om = n~ \epsi^j \om_{j i_2...i_n A_1...A_m}
B^{i_2} \we ...
B^{i_n} \we \mu^{A_1}
\we ... \mu^{A_m} \label{newcontraction}
\en
\noi where $\epsi^j$ {\sl is a 1-form}. This operator still maps
$p$-forms into
$(p-1)$-forms.
We can also define the contraction $i_{\tv_j}$, mapping p-forms into
$(p-2)$-forms,
from
\eq
i_{\epsi^j \tv_j}=\epsi^j i_{\tv_j}
\en
\noi In particular
\eq
 i_{\tv_j} (B^i)=\de^i_j
\en
\noi so that $\tv_j$ can be seen as the ``tangent vector" dual to
$B^j$.
Note that $i_{\epsi^j \tv_j}$ vanishes on $p$-forms that do not
contain at
least
one factor $B^i$.
\sk
ii) a new Lie derivative given by:
\eq
\ell_{\epsi^i \tv_i} \equiv i_{\epsi^i \tv_i}d + d~ i_{\epsi^i \tv_i}
\label{newLie}
\en
This new derivative commutes with $d$, satisfies the Leibnitz rule,
 and acts on the fundamental fields
as
\eqa
& &\ell_{\epsi^j \tv_j}\mu^A = \epsi^j R^A_{~j} \label{newLieonmu}\\
& &\ell_{\epsi^j \tv_j}B^i = d\epsi^i + (C^i_{~Aj} - R^i_{~Aj})
\mu^A \we \epsi^j \label{newLieonB}
\ena

Using these new objects,  and the Bianchi identities
(\ref{Bianchi1}),
(\ref{Bianchi2}),
we can compute the commutator of
two ``usual" Lie derivatives (acting on $\mu^A$ or on $B^i$) and
find:
\eq
\left[ \ell_{ \epsi^A_1 t_A},\ell_{ \epsi^B_2 t_B} \right] = \ell_{
\left[
\epsi^A_1 \partial_A \epsi^C_2 - \epsi^A_2 \partial_A \epsi^C_1
+\epsi^A_1 \epsi^B_2 \left( C^C_{AB} -2R^C_{AB}\right) \right]
\tbo_C} +
 \ell_{\left(C^i_{ABC} - 6R^i_{ABC}\right)\mu^A\epsi^B_1\epsi^C_2
\tbo_i }
\label{FDAalgebra1}
\en
\noi This result has an important consequence: if the field theory
based on FDA1 is geometric, i.e. its action is invariant under
diffeomorphisms
generated by the ``usual" Lie derivative, then the new Lie derivative
defined in (\ref{newLie}) must also generate a symmetry of the
action,
since it appears on the right-hand side of (\ref{FDAalgebra1}).
Thus, when we construct geometric lagrangians gauging FDA1,
we know  {\sl a priori } that the resulting theory will have
symmetries
generated by the new Lie derivative. In other words,  the
transformations
(\ref{newLieonmu}), (\ref{newLieonB}) are invariances of the action.

It becomes now essential to find the total FDA algebra of
transformations,
and show that it closes.
Using the Bianchi identities, it is a straightforward exercise to
find
the remaining commutators:
\eqa
& &\left[ \ell_{ \epsi^A \tbo_A} , \ell_{ \epsi^j \tbo_j} \right] =
\ell_{[
\ell_{ \epsi^A \tbo_A} \epsi^k + \left( C^k_{Bj} - R^k_{Bj} \right)
\epsi^B \epsi^j ] \tbo_k}\\
& &\left[ \ell_{ \epsi^i_1 \tbo_i} , \ell_{ \epsi^j_2 \tbo_j} \right]
=
\ell_{R^B_{~i} (\epsi^i_1 (\epsi_2)^j_B - \epsi_2^i (\epsi_1)^j_B)
\tbo_j}
\ena
\noi where $\epsi^i_A$ are the components of the
1-form $\epsi^i$, i.e.  $\epsi^i \equiv \epsi^i_A \mu^A$.
In particular, we can find the commutators of the Lie derivatives
on the rigid FDA1 ``manifold"  by taking  $\epsi^A =$ const.,
$\epsi^i_B =$
const.
and vanishing curvatures:
\eqa
& &[ \ell_{\tv_A}, \ell_{\tv_B}] = \c{AB}{C} \ell_{\tv_C} +
\cl{ABC}{i} ~
\ell_{\sigma^C \tv_i}\\
& &[\ell_{\tv_A}, \ell_{\si^B \tbo_i}] = [\c{Aj}{k} \de^B_F -
(\c{AF}{B}-2
R^B_{~AF})\de^k_j]
\ell_{\si^F \tbo_k}\\
& &[\ell_{\si^A \tbo_i},\ell_{\si^B \tbo_j}]=0
\ena
This algebra can be considered the dual of the FDA1 system given in
(\ref{fdaLie}), (\ref{fdaB}), and generalizes the
Lie algebra of ordinary Lie derivatives (generating usual
diffeomorphisms)
of (\ref{rigidLie}). Notice the essential presence of
the 1-form $\si$ in front of the ``tangent vectors" $\tbo_i$.
There are  $dim(G) \times dim(D)$ independent extended Lie
derivatives
$\ell_{\si^A \tbo_i}$, with $dim(G)$=dimension of the Lie algebra,
$dim(D)$= dimension of the $D^i_{~j}$ irrep of $G$ to which $B^i$
belongs .

\vfill\eject
\end{document}